\documentclass{elsart}
\usepackage[normalem]{ulem}% Pour distinguer les contribs
\usepackage{graphicx}
\usepackage{latexsym,amssymb,amsmath,amssymb,xspace,theorem}
\usepackage{multicol}
\usepackage{t1enc}
\usepackage{epic,eepic}
\usepackage{subfigure}
\usepackage{epsfig}
\theoremstyle{plain} \theoremheaderfont{\scshape}

\newtheorem{Thm}{\bf Theorem}
\newtheorem{Lem}[Thm]{\bf Lemma}
\newtheorem{Clm}{Claim}[Thm]

\newtheorem{Conj}[Thm]{{\bf Conjecture}}

\newtheorem{Prop}[Thm]{\bf Proposition}
\newtheorem{Cor}[Thm]{ \bf Corollary}
{\theorembodyfont{\rmfamily}
 \newtheorem{Def}[Thm]{\bf Definition}
 \newtheorem{Rem}[Thm]{\bf Remark}

}

\newenvironment{Prf}{{\bf \noindent Proof } }{\hfill$\square$\\}
\newenvironment{PrfClaim}{{\bf Proof }}{{\hfill\tiny{$\square$\\}}}

\newcommand{\ignore}[1]{}

\newcommand{\cqfd}{\unskip\kern 6pt\penalty 500
\raise -2pt\hbox{\vrule\vbox to 10pt{\hrule width 4pt
\vfill\hrule}\vrule}\par}

\newcommand{\cubthree}{cubic $3$-edge colourable graph \xspace}

\newcommand{\cubthreev}{cubic $3$-edge colourable graph}
\newcommand{\cubthreesv}{cubic $3$-edge colourable graphs}

\newcommand{\Jae}{Jaeger's graph \xspace}
\newcommand{\Jaes}{Jaeger's graphs \xspace}
\newcommand{\Jaev}{Jaeger's graph}
\newcommand{\Jaesv}{Jaeger's graphs}

\input{epsf.sty}
\begin{document}
\begin{frontmatter}


\title{On a family of cubic graphs containing the flower snarks}
\author{Jean-Luc Fouquet \and Henri Thuillier \and Jean-Marie Vanherpe}
\address{L.I.F.O., Facult\'e des Sciences, B.P. 6759 \\ Universit\'e d'Orl\'eans, 45067 Orl\'eans Cedex 2, FR}
\begin{abstract}
We consider cubic graphs formed with $k \geq 2$ disjoint claws $C_i
\sim K_{1, 3}$ ($0 \leq i \leq k-1$) such that for every integer $i$
modulo $k$ the three vertices of degree $1$ of $\ C_i$ are joined to
the three vertices of degree $1$ of $C_{i-1}$ and joined to the
three vertices of degree $1$ of $C_{i+1}$. Denote by $t_i$ the
vertex of degree $3$ of $C_i$ and by $T$ the set $\{t_1, t_2,...,
t_{k-1}\}$. In such a way we construct three distinct graphs, namely
$FS(1,k)$, $FS(2,k)$ and $FS(3,k)$. The graph $FS(j,k)$ ($j \in \{1,
2, 3\}$) is the graph where the set of vertices
$\cup_{i=0}^{i=k-1}V(C_i) \setminus T$ induce $j$ cycles (note that
the graphs $FS(2,2p+1)$, $p\geq2$, are the flower snarks defined by
Isaacs \cite{Isa75}). We determine the number of perfect matchings
of every $FS(j,k)$. A cubic graph $G$ is said to be {\em $2$-factor
hamiltonian} if every $2$-factor of $G$ is a hamiltonian cycle. We
characterize the graphs $FS(j,k)$ that are $2$-factor hamiltonian
(note that $FS(1,3)$ is the "Triplex Graph" of Robertson, Seymour
and Thomas \cite{RobSey}). A {\em strong matching} $M$ in a graph
$G$ is a matching $M$ such that there is no edge of $E(G)$
connecting any two edges of $M$. A cubic graph having a perfect
matching union of two strong matchings is said to be a {\em\Jaev}.
We characterize the graphs $FS(j,k)$ that are \Jaesv.
\end{abstract}

\begin{keyword}
cubic graph;  perfect matching; strong matching; counting;
hamiltonian cycle; $2$-factor hamiltonian
\end{keyword}
\end{frontmatter}
%\maketitle

\section{Introduction}

The complete bipartite graph $K_{1,3}$ is called, as usually, a {\em
claw}. Let $k$ be an integer $\geq 2$ and let $G$ be a cubic graph
on $4k$ vertices formed with $k$ disjoint claws
$C_i=\{x_i,y_i,z_i,t_i\}$ ($0 \leq i \leq k-1$) where $t_i$ (the
{\em center} of $C_{i}$) is joined to the three independent vertices
$x_i,y_i$ and $z_i$ (the {\em external} vertices of $C_{i}$). For
every integer $i$ modulo $k$ $\ C_i$ has three neighbours in
$C_{i-1}$ and three neighbours in $C_{i+1}$. For any integer $k \geq
2$ we shall denote the set of integers modulo $k$ as $ {\bf
Z}_{k}^{}$. In the sequel of this paper indices $i$ of claws $C_i$
belong to $ {\bf Z}_{k}^{}$.

\begin{figure}[htb]
\centering \epsfsize=0.5 \hsize \noindent \epsfbox{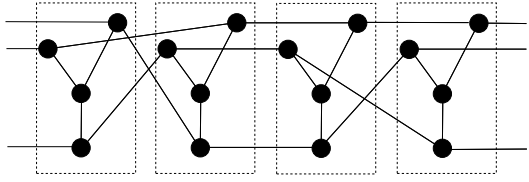}
\caption{Four consecutive claws} \label{Figure:fsk}
\end{figure}

By renaming some external vertices of claws we can suppose, without
loss of generality,  that $\{x_i x_{i+1},  y_i y_{i+1}, z_i
z_{i+1}\}$ are edges for any $i$ distinct from $k-1$. That is to say
the subgraph induced on $X=\{x_0, x_1, \ldots, x_{k-1} \}$
(respectively $Y=\{y_0, y_1, \ldots, y_{k-1} \}$, $Z=\{z_0, z_1,
\ldots, z_{k-1} \}$) is a path or a cycle (as induced subgraph of
$G$). Denote by $T$ the set of the internal vertices $\{t_0, t_1,
\ldots, t_{k-1} \}$.

Up to isomorphism, the matching joining the external vertices of
$C_{k-1}$ to those of $C_{0}$ (also called, for $k \geq 3$, {\em
edges between $C_{k-1}$ and $C_{0}$}) determines the graph G. In
this way we construct essentially three distinct graphs, namely
$FS(1,k)$, $FS(2,k)$ and $FS(3,k)$. The graph $FS(j,k)$ ($j \in \{1,
2, 3\}$) is the graph where the set of  vertices
$\cup_{i=0}^{i=k-1}\{C_i \setminus \{t_i\} \}$ induces $j$ cycles.
For $k \geq 3$ and any $j \in \{1, 2, 3\}$ the graph $FS(j, k)$ is a
simple cubic graph. When $k$ is odd, the $FS(2,k)$ are the graphs
known as the flower snarks \cite{Isa75}. We note that $FS(3, 2)$ and
$FS(2, 2)$ are multigraphs, and that $FS(1, 2)$ is isomorphic to the
cube. For $k=2$ the notion of "edge between $C_{k-1}$ and $C_{0}$"
is ambiguous, so we must define it precisely. For two parallel edges
having one end in $C_0$ and the other in $C_1$, for instance two
parallel edges having $x_0$ and $x_1$ as endvertices, we denote one
edge by $x_0x_1$ and the other by $x_1x_0$. An edge in $\ \{x_1x_0,
x_1y_0, x_1z_0, y_1x_0, y_1y_0, y_1z_0, z_1x_0, z_1y_0, z_1z_0\}$,
if it exists, is an {\em edge between $C_1$ and $C_0$}. We will say
that $x_0x_1$, $y_0y_1$ and $z_0z_1$ are {\em edges between $C_0$
and $C_1$}.

By using an ad hoc translation of the indices of claws (and of their
vertices) and renaming some external vertices of claws, we see that
for any reasoning about a sequence of $h \geq 3$ consecutive claws
$(C_i, C_{i+1}, C_{i+2}, \ldots, C_{i+h-1})$ there is no loss of
generality to suppose that $0 \leq i < i+h-1 \leq k-1$. For a
sequence of claws $(C_p,..., C_r)$ with $0 \leq p < r \leq k-1$,
since $0$ is a possible value for subscript $p$ and since $k-1$ is a
possible value for subscript $r$, it will be useful from time to
time to denote by $x'_{p-1}$ the neighbour in $C_{p-1}$ of the
vertex $x_p$ of $C_{p}$ (recall that $x'_{p-1}\in
\{x_{k-1},y_{k-1},z_{k-1}\}$ if $p=0$), and to denote by $x'_{r+1}$
the neighbour in $C_{r+1}$ of the vertex $x_{r}$ of $C_{r}$ (recall
that $x'_{r+1}\in \{x_{0},y_{0},z_{0}\}$ if $r=k-1$). We shall make
use of analogous notations for neighbours of $y_p$, $z_p$, $y_r$ and
$z_r$.

We shall prove in the following lemma that there are essentially two
types of perfect matchings in $FS(j,k)$.

\begin{Lem} \label{Lemma:PerfectMatching} Let $G\in \{ FS(j,k),
j \in \{1, 2, 3\}, k \geq 2$ and let $M$ be a perfect matching of
$G$. Then the $2-$factor $G \setminus M$ induces a path of length
$2$ and an isolated vertex in each claw $C_{i}$  ($i \in {\bf
Z}_{k}^{}$) and $M$ fulfils one (and only one) of the three
following properties :
\begin{itemize}
  \item[i)] For every $i$ in ${\bf Z}_{k}^{}$ $ M$ contains exactly one edge joining the claw $C_i$ to the claw
$C_{i+1}$,
  \item[ii)] For every even $i$ in ${\bf Z}_{k}^{}$ $\ M$ contains exactly two edges between $C_i$ and $C_{i+1}$ and
none between $C_{i-1}$ and $C_i$,
  \item[iii)] For every odd $i$ in ${\bf Z}_{k}^{}$ $\ M$ contains exactly two edges between $C_i$ and $C_{i+1}$ and
none between $C_{i-1}$ and $C_i$.
\end{itemize}
Moreover, when $k$ is odd $\ M$ satisfies only item {\it i)}.
\end{Lem}

\begin{Prf}
Let $M$ be a perfect matching of $G = FS(j,k)$ for some $j \in \{1,
2, 3\}$. Since $M$ contains exactly one edge of each claw, it is
obvious that $G \setminus M$ induces a path of length $2$ and an
isolated vertex in each claw $C_{i}$.

For each claw $C_{i}$ of $G$ the vertex $t_{i}$ must be saturated by
an edge of $M$ whose end (distinct from $t_{i}$) is in
$\{x_{i},y_{i},z_{i}\}$. Hence there are exactly two edges of $M$
having one end in $C_{i}$ and the other in $C_{i-1} \cup C_{i+1}$.

If there are two edges of $M$ between $C_{i}$ and $C_{i+1}$ then
there is no edge of $M$ between $C_{i-1}$ and $C_{i}$. If there are
two edges of $M$ between $C_{i-1}$ and $C_{i}$ then there is no edge
of $M$ between $C_{i}$ and $C_{i+1}$. Hence, we get {\it ii)} or
{\it iii)} and we must have an even number $k$ of claws in $G$.

Assume now that there is only one edge of $M$ between $C_{i-1}$ and
$C_{i}$. Then there exists exactly one edge between $C_{i}$ and
$C_{i+1}$ and, extending this trick to each claw of $G$, we get {\it
i)} when $k$ is even or odd.
\end{Prf}

\begin{Def} \label{Definition_type_of_matching}
We say that a perfect matching $M$ of $FS(j, k)$ is of {\em type
$1$} in Case i) of Lemma \ref{Lemma:PerfectMatching} and of {\em
type $2$} in Cases ii) and iii). If neccessary, to distinguish Case
ii) from Case iii) we shall say {\em type $2.0$} in Case ii) and
{\em type $2.1$} in Case iii). We note that the numbers of perfect
matchings of type $2.0$ and of type $2.1$ are equal.
\end{Def}

\noindent
{\bf Notation} : The length of a path $P$ (respectively a
cycle $\Gamma$) is denoted by $l(P)$ (respectively $l(\Gamma)$).

\section{Counting perfect matchings of $FS(j, k)$}\label{Section:CountingPerfectMatchings}

We shall say that a vertex $v$ of a cubic graph $G$ is {\em
inflated} into a triangle when we construct a new cubic graph $G'$
by deleting $v$ and adding three new vertices inducing a triangle
and joining each vertex of the neighbourhood $N(v)$ of $v$ to a
single vertex of this new triangle. We say also that $G'$ is
obtained from $G$ by a {\em triangular extension}. The converse
operation is the {\em contraction} or {\em reduction} of the
triangle. The number of perfect matchings of $G$ is denoted by
$\mu(G)$.

\begin{Lem} \label{Lemma:InflatingBipartite} Let $G$ be a bipartite
cubic graph and let $\{V_{1},V_{2}\}$ be the bipartition of its
vertex set. Assume that each vertex in some subset $W_{1} \subseteq
V_{1}$ is inflated into a triangle and let $G'$ be the graph
obtained in that way. Then $\mu(G)=\mu(G')$.

\end{Lem}

\begin{Prf}
Note that  $\{V_{1},V_{2}\}$ is a balanced bipartition and, by
K�nig's Theorem, the graph $G$ is a \cubthreev. So, $G'$ is also a
\cubthree (hence, $G$ and $G'$ have perfect matchings). Let $M$ be a
perfect matching of $G'$. Each vertex of $V_{1} \setminus W_{1}$ is
saturated by an edge whose second end vertex is in $V_{2}$. Let $A
\subseteq V_{2}$ be the set of vertices so saturated in $V_{2}$.
Assume that some triangle of $G'$ is such that the three vertices
are saturated by three edges having one end in the triangle and the
second one in $V_{2}$. Then we need to have at least $|W_{1}|+2$
vertices in $V_{2} \setminus A$, a contradiction. Hence, $M$ must
have exactly one edge in each triangle and the contraction of each
triangle in order to get back $G$ transforms $M$ in a perfect
matching of $G$. Conversely, each perfect matching of $G$ leads to a
unique perfect matching of $G'$ and we obtain the result.
\end{Prf}

Let us denote by $\mu(j,k)$ the number of perfect matchings of
$FS(j,k)$, $\mu_{1}(j,k)$ its number of perfect matchings of type
$1$ and $\mu_{2}(j,k)$ its number of perfect matchings of type $2$.

\begin{Lem} \label{Lemma:SmallCases} We have

\begin{itemize}
  \item $\mu(1,3)=\mu_{1}(1,3)=9$
  \item $\mu(2,3)=\mu_{1}(2,3)=8$
  \item $\mu(3,3)=\mu_{1}(3,3)=6$
  \item $\mu(1,2)=9$, $\mu_{1}(1,2)=3$
  \item $\mu(2,2)=10$, $\mu_{1}(2,2)=4$
  \item $\mu(3,2)=12$, $\mu_{1}(3,2)=6$
\end{itemize}

\end{Lem}

\begin{Prf}
The cycle containing the external vertices of the claws of the graph
$FS(1,3)$ is $x_0, x_1, x_2, y_0, y_1, y_2, z_0, z_1, z_2, x_0$.
Consider a perfect matching $M$ containing the edge $t_0x_0$. There
are two cases: $i)$ $x_1x_2 \in M$ and $ii)$ $x_1t_1 \in M$. In Case
$i)$ we must have $y_0y_1, t_1z_1, t_2z_2, z_0y_2 \in M$. In Case
$ii)$ there are two sub-cases: $ii).a$ $x_2y_0 \in M$ and $ii).b$
$x_2t_2 \in M$. In Case $ii).a$ we must have $y_1y_2, t_2z_2, z_0z_1
\in M$ and in Case $ii).b$ we must have $y_0y_1, y_2z_0, z_1z_2 \in
M$. Thus, there are exactly $3$ distinct perfect matching containing
$t_0x_0$. By symmetry, there are $3$ distinct perfect matchings
containing $t_0y_0$, and $3$ distinct matchings containing $t_0z_0$,
therefore $\mu(1,3)=9$.

It is well known that the Petersen graph has exactly $6$ perfect
matchings. Since $FS(2,3)$ is obtained from the Petersen graph  by
inflating a vertex into a triangle these $6$ perfect matchings lead
to $6$ perfect matchings of  $FS(2,3)$. We have two new perfect
matchings when considering the three edges connected to this
triangle (we have two ways to include these edges into a perfect
matching). Hence  $\mu(2,3)=8$.

$FS(3,3)$ is obtained from $K_{3,3}$ by inflating three vertices in
the same colour of the bipartition. Since $K_{3,3}$ has six perfect
matchings, applying Lemma \ref{Lemma:InflatingBipartite} we get
immediately the result for $\mu(3,3)$.

Is is a routine matter to obtain the values for $FS(j,2)\ $ ($j \in
\{1, 2, 3 \}$).
\end{Prf}

\begin{Thm} The numbers $\mu(i,k)$ of perfect matchings of $FS(i,k)$ ($i \in \{1,2,3\}$)
are given by:

When $k$ is odd
\begin{tabular}{c l }
  $\bullet$ & $\mu(2,k)= 2^k$  \\
  $\bullet$ & $\mu(1,k)= 2^{k}+1$  \\
  $\bullet$ & $\mu(3,k)= 2^{k}-2$ \\

\end{tabular}

When $k$ is even
\begin{tabular}{c l }
 $\bullet$ & $\mu(2,k)= 2 \times 3^{\frac{k}{2}}+2^{k}$  \\
 $\bullet$ & $\mu(1,k)= 2 \times 3^{\frac{k}{2}}+2^{k}-1$  \\
   $\bullet$ & $\mu(3,k)= 2 \times 3^{\frac{k}{2}}+2^{k}+2$  \\
\end{tabular}

\end{Thm}
\begin{Prf}
We shall prove this result by induction on $k$ and we distinguish
the case "$k$ odd"  and the case "$k$ even".

The following trick will be helpful. Let $i \not = 0$ and let
$C_{i-2}$, $C_{i-1}$, $C_{i}$ and $C_{i+1}$ be four consecutive
claws of $FS(j,k)$ ($j \in \{1, 2, 3\}$). We can delete $C_{i-1}$
and $C_{i}$ and join the three external vertices of $C_{i-2}$ to the
three external vertices of $C_{i+1}$ by a matching in such a way
that the resulting graph is $FS(j',k-2)$. We have three distinct
ways to reduce $FS(j,k)$ into $FS(j',k-2)$ when deleting $C_{i-1}$
and $C_{i}$.

{\bf Case 1:} We add the edges $\{x_{i-2}x_{i+1},
y_{i-2}y_{i+1},z_{i-2}z_{i+1}\}$ and we get $G_{1}=FS(j_{1},k-2)$

{\bf Case 2:} We add the edges $\{x_{i-2}y_{i+1},
y_{i-2}z_{i+1},z_{i-2}x_{i+1}\}$ and we get  $G_{2}=FS(j_{2},k-2)$.

{\bf Case 3:} We add the edges $\{x_{i-2}z_{i+1},
y_{i-2}x_{i+1},z_{i-2}y_{i+1}\}$ and we get $G_{3}=FS(j_{3},k-2)$.

Following the cases, we shall precise the values of $j_{1},j_{2}$
and $j_{3}$.

It is an easy task to see that each perfect matching of type $1$ of
$FS(j,k)$ leads to a perfect matching of either $G_{1}$ or $G_{2}$
or $G_{3}$ and, conversely, each perfect matching of type $1$ of
$G_{1}$ allows us to construct $2$ distinct perfect matchings of
type $1$ of $FS(j,k)$, while each perfect matching of type $1$ of
$G_{2}$ and $G_{3}$ allows us to construct $1$ perfect matching of
type $1$ of $FS(j,k)$.

  We have

\begin{equation}\label{Equation:1}
    \mu_{1}(j,k)=2\mu_{1}(G_{1})+\mu_{1}(G_{2})+\mu_{1}(G_{3})
\end{equation}

\begin{Clm} \label{Claim:mu2k}$\mu_{1}(2,k)=2^{k}$
\end{Clm}
\begin{PrfClaim}
Since the result holds  for $FS(2,3)$ and $FS(2,2)$ by Lemma
\ref{Lemma:SmallCases}, in order to prove the result by induction on
the number $k$ of claws, we assume that the property holds  for
$FS(2,k-2)$ with $k-2 \geq 2$.

In that case $G_{1}$, $G_{2}$ and $G_{3}$ are isomorphic to
$FS(2,k-2)$.   Using Equation \ref{Equation:1} we have, as claimed

$$\mu_{1}(2,k)=4\mu_{1}(2,k-2)=2^{k}$$

\end{PrfClaim}

\begin{Clm}\label{Claim:mu1k3k}
$\mu_{1}(1,k)=2^{k}-(-1)^k$ and $\mu_{1}(3,k)=2^{k}+2(-1)^k$
\end{Clm}
\begin{PrfClaim}
Since the result holds  for $FS(1,3)$, $FS(1,2)$, $FS(3,3)$, and
$FS(3,2)$, by Lemma \ref{Lemma:SmallCases}, in order to prove the
result by induction on the number $k$ of claws, we assume that the
property holds  for $FS(1,k-2)$, and $FS(3,k-2)$ with $k-2 \geq 2$.

When considering $FS(1,k)$, $G_{1}$ is isomorphic to $FS(1,k-2)$,
and among $G_{2}$ and $G_{3}$ one of them  is isomorphic to
$FS(3,k-2)$ and the other to $FS(1,k-2)$. In the same way, when
considering $FS(3,k)$, $G_{1}$ is isomorphic to $FS(3,k-2)$, and
 $G_{2}$ and $G_{3}$ are isomorphic to $FS(1,k-2)$.

Using Equation \ref{Equation:1} we have,
$$\mu_{1}(1,k)=2\mu_{1}(1,k-2)+\mu_{1}(1,k-2)+\mu_{1}(3,k-2)$$
and
$$\mu_{1}(1,k)=2(2^{k-2}+1)+2^{k-2}+1+2^{k-2}-2=2^{k}+1$$
$$\mu_{1}(3,k)=2(2^{k-2}-2)+2^{k-2}+1+2^{k-2}+1=2^{k}-2$$

\end{PrfClaim}

When $k$ is odd, we have $\mu_{2}(j,k)=0$ by Lemma
\ref{Lemma:PerfectMatching} and hence $\mu(j,k)=\mu_{1}(j,k)$

 When $k$ is even it remains to count the number of perfect matchings of type
$2$. From Lemma \ref{Lemma:PerfectMatching}, for every two
consecutive claws $C_{i}$ and $C_{i+1}$, we have either two edges of
$M$ joining the external vertices of $C_{i}$ to those of $C_{i+1}$
or none. We have $3$ ways to choose $2$ edges between $C_{i}$ and
$C_{i+1}$, each choice of these two edges can be completed in a
unique way in a perfect matching of the subgraph $C_{i} \cup
C_{i+1}$. Hence we get easily that the number of perfect matchings
of type $2$ in $FS(j,k)$ ($j \in \{1, 2, 3\}$) is
\begin{equation}\label{Equation:mu2}
     \mu_{2}(j,k)=2 \times 3^{\frac{k}{2}}
\end{equation}

Using Claims \ref{Claim:mu2k} and \ref{Claim:mu1k3k} and Equation
\ref{Equation:mu2} we get the results for $\mu(j,k)$ when $k$ is
even.

\end{Prf}

\section{Some structural results about perfect matchings of FS(j, k)}

\subsection{Perfect matchings of type 1}
\label{subsect:type-1}

%We develop here some structural results about perfect matchings of
%type 1 in $\ G=FS(j,k)$.

\begin{Lem} \label{Lemma:2_factor_type1_1}
Let $M$ be a perfect matching of type $1$ of $\ G=FS(j,k)$. Then the
$2-$factor $G \setminus M$ has exactly one or two cycles and each
cycle of $\ G \setminus M$ has at least one vertex in each claw
$C_{i}$ ($i \in {\bf Z}_{k}^{}$).
\end{Lem}

\begin{Prf} Let $M$ be a perfect matching of type $1$ in
$G$. Let us consider the claw $C_{i}$ for some $i$ in ${\bf
Z}_{k}^{}$. Assume without loss of generality that the edge of $M$
contained in $C_{i}$ is $t_{i}x_{i}$. The cycle of $G \setminus M$
visiting $x_{i}$ comes from $C_{i-1}$, crosses $C_{i}$ by using the
vertex $x_{i}$ and goes to $C_{i+1}$. By Lemma
\ref{Lemma:PerfectMatching}, the path $y_{i}t_{i}z_{i}$ is contained
in a cycle of $G \setminus M$. The two edges incident to $y_{i}$ and
$z_{i}$ joining $C_{i}$ to $C_{i-1}$ (as well as those joining
$C_{i}$ to $C_{i+1}$) are not contained both in $M$ (since $M$ has
type $1$). Thus, the cycle of $G \setminus M$ containing
$y_{i}t_{i}z_{i}$ comes from $C_{i-1}$, crosses $C_{i}$  and goes to
$C_{i+1}$. Thus, we have at most two cycles in $G \setminus M$, as
claimed, and we can note that each claw must be visited by these
cycles.
\end{Prf}

\begin{Def}
Let us suppose that $M$ is a perfect matching of type 1 in $G =
FS(j, k)$ such that the $2-$factor $G \setminus M$ has exactly two
cycles $\Gamma_1$ and $\Gamma_2$. A claw $C_i$ intersected by three
vertices of $\Gamma_{1}$ (respectively $\Gamma_{2}$) is said to be
{\em $\Gamma_{1}$-major} (respectively {\em $\Gamma_{2}$-major)}.
\end{Def}

\begin{Lem} \label{Lemma:2_factor_type1_2}
Let $M$ be a perfect matching of type $1$ of $\ G=FS(j,k)$ such that
the $2-$factor $G \setminus M$ has exactly two cycles. Then, the
lengths of these two cycles have the same parity as $k$, and those
lengths are distinct when $k$ is odd.
\end{Lem}

\begin{Prf} Let $\Gamma_{1}$ and $\Gamma_{2}$ be the two cycles of $G
\setminus M$. By Lemma \ref{Lemma:2_factor_type1_1}, for each $i$ in
${\bf Z}_{k}^{}$ these two cycles must cross the claw $C_{i}$. Let
$k_{1}$ be the number of $\Gamma_{1}$-major claws and let $k_{2}$ be
the number of $\Gamma_{2}$-major claws. We have $k_{1}+k_{2}=k$,
$l(\Gamma_{1})= 3k_1+k_2$ and $l(\Gamma_{2})= 3k_{2}+k_{1}$. When
$k$ is odd, we must have either $k_{1}$ odd and $k_{2}$ even, or
$k_{1}$ even and $k_{2}$ odd. Then $\Gamma_{1}$ and $\Gamma_{2}$
have distinct odd lengths. When $k$ is even, we must have either
$k_{1}$ and $k_{2}$ even, or $k_{2}$ and $k_{1}$ odd. Then
$\Gamma_{1}$ and $\Gamma_{2}$ have even lengths.
\end{Prf}

\begin{Lem} \label{Lemma:Gamma1-Gamma1}
Let $M$ be a perfect matching of type $1$ of $\ G=FS(j,k)$ such that
the $2-$factor $G \setminus M$ has exactly two cycles $\Gamma_1$ and
$\Gamma_2$. Suppose that there are two consecutive $\Gamma_1$-major
claws $C_j$ and $C_{j+1}$ with $j \in {\bf Z}_{k}^{} \setminus
\{k-1\}$. Then there is a perfect matching $M'$ of type $1$ such
that the $2-$factor $G \setminus M'$ has exactly two cycles
$\Gamma'_1$ and $\Gamma'_2$ having the following properties:
\begin{itemize}
 \item[a)] for $i \in {\bf Z}_{k}^{} \setminus \{j,j+1\}$ $C_i$ is
$\Gamma'_2$-major if and only if $C_i$ is $\Gamma_2$-major,
 \item[b)] $C_j$and $C_{j+1}$ are $\Gamma'_2$-major,
 \item[c)] $l(\Gamma'_1) = l(\Gamma_1)-4$ and $\ l(\Gamma'_2) =
l(\Gamma_2)+4$.
\end{itemize}
\end{Lem}

\begin{figure}[htb]
\centering \epsfsize=0.8 \hsize \noindent
\epsfbox{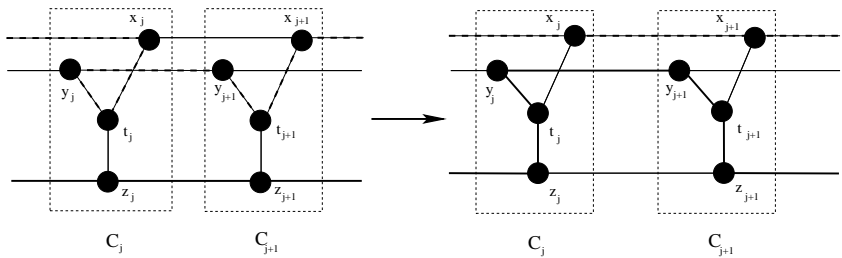} \caption{Local transformation
of type 1} \label{Figure:local-transformation-1}
\end{figure}

\begin{Prf}
Consider the claws $C_j$ and $C_{j+1}$. Since $C_j$ is a
$\Gamma_1$-major claw suppose without loss of generality that
$t_jz_j$ belongs to $M$ and that $\Gamma_1$ contains the path
$x'_{j-1}x_jt_jy_jy_{j+1}$  where $x'_{j-1}$ denotes the neighbour
of $x_j$ in $C_{j-1}$ (then $x_jx_{j+1}$ belongs to $M$). Since
$C_{j+1}$ is $\Gamma_1$-major and $\Gamma_2$ goes through $C_j$ and
$C_{j+1}$, the cycle $\Gamma_1$ must contain the path
$y_{j+1}t_{j+1}x_{j+1}x'_{j+2}$ where $x'_{j+2}$ denotes the
neighbour of $x_{j+1}$ in $C_{j+2}$ (then $M$ contains
$t_{j+1}z_{j+1}$ and $y_{j+1}y'_{j+2}$). Denote by $P_1$ the path
$x'_{j-1}x_jt_jy_jy_{j+1}t_{j+1}x_{j+1}x'_{j+2}$. Note that
$\Gamma_2$ contains the path $P_2 = z'_{j-1}z_jz_{j+1}z'_{j+1}$
where $z'_{j-1}$ and $z'_{j+1}$ are defined similarly.  See to the
left part of Figure \ref{Figure:local-transformation-1}.

Let us perform the following local transformation: delete
$x_jx_{j+1}$, $t_jz_j$ and $t_{j+1}z_{j+1}$ from $M$ and add
$z_jz_{j+1}$, $t_jx_j$ and $t_{j+1}x_{j+1}$. Let $M'$ be the
resulting perfect matching. Then the subpath $P_1$ of $\Gamma_1$ is
replaced by $P'_1=x'_{j-1}x_jx_{j+1}x'_{j+2}$ and the subpath $P_2$
of $\Gamma_2$ is replaced by
$P'_2=z'_{j-1}z_jt_jy_jy_{j+1}t_{j+1}z_{j+1}z'_{j+2}$ (see Figure
\ref{Figure:local-transformation-1}). We obtain a new $2$-factor
containing two new cycles $\Gamma'_{1}$ and $\Gamma'_{2}$. Note that
$C_j$ and $C_{j+1}$ are $\Gamma'_2$-major claws and for $i$ in ${\bf
Z}_{k}^{} \setminus \{j,j+1\}$ $C_i$ is $\Gamma'_2$-major
(respectively $\Gamma'_1$-major) if and only if $C_i$ is
$\Gamma_2$-major (respectively $\Gamma_1$-major). The length of
$\Gamma_{1}$ (now $\Gamma'_{1}$) decreases of $4$ units while the
length of $\Gamma_{2}$ (now $\Gamma'_{2}$) increases of $4$ units.
\end{Prf}

\vspace{5mm}

The operation depicted in Lemma \ref{Lemma:Gamma1-Gamma1} above will
be called a {\em local transformation of type 1}.

\begin{Lem} \label{Lemma:Gamma1-Gamma2-Gamma1}
Let $M$ be a perfect matching of type $1$ of $\ G=FS(j,k)$ such that
the $2-$factor $G \setminus M$ has exactly two cycles $\Gamma_1$ and
$\Gamma_2$. Suppose that there are three consecutive claws $C_j$,
$C_{j+1}$ and $C_{j+2}$ with $j$ in ${\bf Z}_{k}^{} \setminus \{k-1,
k-2\}$ such that $C_j$ and $C_{j+2}$ are $\Gamma_1$-major and
$C_{j+1}$ is $\Gamma_2$-major. Then there is a perfect matching $M'$
of type $1$ such that the $2-$factor $G \setminus M'$ has exactly
two cycles $\Gamma'_1$ and $\Gamma'_2$ having the following
properties:
\begin{itemize}
 \item[a)] for $i \in {\bf Z}_{k}^{} \setminus \{j,j+1,j+2\}$ $C_i$ is
$\Gamma'_2$-major if and only if $C_i$ is $\Gamma_2$-major,
 \item[b)] $C_j$ and $C_{j+2}$ are $\Gamma'_2$-major and $C_{j+1}$
is $\Gamma'_1$-major,
 \item[c)]  $l(\Gamma'_1) = l(\Gamma_1)-2$ and $\ l(\Gamma'_2) =
l(\Gamma_2)+2$.
\end{itemize}
\end{Lem}

\begin{figure}[htb]
\centering \epsfsize=0.999 \hsize \noindent
\epsfbox{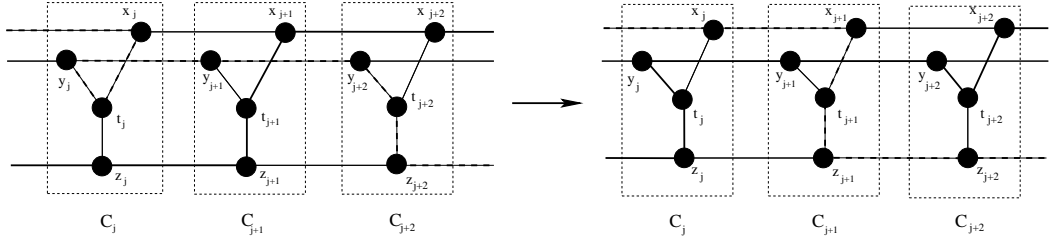} \caption{Local transformation
of type 2} \label{Figure:local-transformation-2}
\end{figure}

\begin{Prf}
Since $C_{j}$ is $\Gamma_1$-major, as in the proof of Lemma
\ref{Lemma:Gamma1-Gamma1} suppose that $\Gamma_1$ contains the path
$x'_{j-1}x_jt_jy_jy_{j+1}$ (that is edges $t_jz_{j}$ and
$x_{j}x_{j+1}$ belong to $M$). Since $C_{j+1}$ is $\Gamma_2$-major
the cycle $\Gamma_1$ contains the edge $y_{j+1}y_{j+2}$. Then we see
that $\Gamma_1$ contains the path
$Q_1=x'_{j-1}x_jt_jy_jy_{j+1}y_{j+2}t_{j+2}z_{j+2}z'_{j+3}$ and that
$\Gamma_2$ contains the path
$Q_2=z'_{j-1}z_jz_{j+1}t_{j+1}x_{j+1}x_{j+2}x'_{j+3}$. Note that
$y_{j+1}t_{j+1}$, $z_{j+1}z_{j+2}$ and $t_{j+2}x_{j+2}$ belong to
$M$.

Let us perform the following local transformation: delete
$t_jz_{j}$, $x_jx_{j+1}$, $z_{j+1}z_{j+2}$ and $x_{j+2}t_{j+2}$ from
$M$ and add $x_jt_j$, $z_{j}z_{j+1}$, $x_{j+1}x_{j+2}$ and
$z_{j+2}t_{j+2}$ to $M$. Let $M'$ be the resulting perfect matching.
Then the subpath $Q_1$ of $\Gamma_1$ is replaced by
$Q'_1=x'_{j-1}x_jx_{j+1}t_{j+1}z_{j+1}z_{j+2}z'_{j+3}$ and the
subpath $Q_2$ of $\Gamma_2$ is replaced by
$Q'_2=z'_{j-1}z_jt_jy_jy_{j+1}y_{j+2}t_{j+2}x_{j+2}x'_{j+3}$ (see
Figure \ref{Figure:local-transformation-2}). We obtain a new
$2$-factor containing two new cycles named $\Gamma'_{1}$ and
$\Gamma'_{2}$. Note that $C_j$ and $C_{j+2}$ are now
$\Gamma'_2$-major claws and $C_{j+1}$ is $\Gamma'_1$-major. The
length of $\Gamma_{1}$ decreases of $2$ units while the length of
$\Gamma_{2}$ increases of $2$ units. It is clear that for $i \in
{\bf Z}_{k}^{} \setminus \{j,j+1,j+2\}$ $C_i$ is $\Gamma'_2$-major
(respectively $\Gamma'_1$-major) if and only if $C_i$ is
$\Gamma_2$-major (respectively $\Gamma_1$-major).
\end{Prf}

\vspace{1cm}

The operation depicted in Lemma \ref{Lemma:Gamma1-Gamma2-Gamma1}
above will be called a {\em local transformation of type 2}.

\begin{Lem} \label{Lemma:Gamma1-Gamma2-Gamma2}
Let $M$ be a perfect matching of type $1$ of $\ G=FS(j,k)$ such that
the $2-$factor $G \setminus M$ has exactly two cycles $\Gamma_1$ and
$\Gamma_2$. Suppose that there are three consecutive claws $C_j$,
$C_{j+1}$ and $C_{j+2}$ with $j$ in ${\bf Z}_{k}^{} \setminus \{k-1,
k-2\}$ such that $C_{j+1}$ and $C_{j+2}$ are $\Gamma_2$-major and
$C_j$ is $\Gamma_1$-major. Then there is a perfect matching $M'$ of
type $1$ such that the $2-$factor $G \setminus M'$ has exactly two
cycles $\Gamma'_1$ and $\Gamma'_2$ having the following properties:
\begin{itemize}
 \item[a)]  for $i \in {\bf Z}_{k}^{} \setminus \{j,j+1,j+2\}$
$C_i$ is $\Gamma'_2$-major if and only if $C_i$ is $\Gamma_2$-major,
 \item[b)]   $C_j$ and $C_{j+1}$ are $\Gamma'_2$-major and $C_{j+2}$
is $\Gamma'_1$-major,
 \item[c)]   $l(\Gamma'_1) = l(\Gamma_1)$ and $\ l(\Gamma'_2) =
l(\Gamma_2)$.
\end{itemize}
\end{Lem}

\begin{figure}[htb]
\centering \epsfsize=0.999 \hsize \noindent
\epsfbox{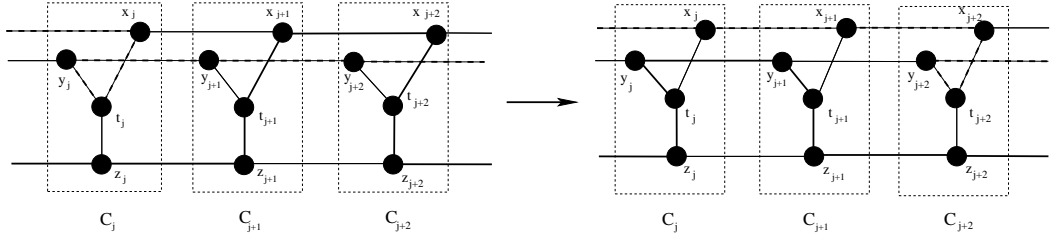} \caption{Local transformation
of type 3} \label{Figure:local-transformation-3}
\end{figure}

\begin{Prf}
Since $C_{j}$ is $\Gamma_1$-major, as in the proof of Lemma
\ref{Lemma:Gamma1-Gamma1} suppose that $\Gamma_1$ contains the path
$x'_{j-1}x_jt_jy_jy_{j+1}$ (that is edges $t_jz_{j}$ and
$x_{j}x_{j+1}$ belong to $M$). Since $C_{j+1}$ and $C_{j+2}$ are
$\Gamma_2$-major, the unique vertex of $C_{j+1}$ (respectively
$C_{j+2}$) contained in $\Gamma_1$ is $y_{j+1}$ (respectively
$y_{j+2}$). Note that the perfect matching $M$ contains the edges
$t_jz_j$, $x_jx_{j+1}$, $t_{j+1}y_{j+1}$, $z_{j+1}z_{j+2}$ and
$t_{j+2}y_{j+2}$. Then the path $R_1 =
x'_{j-1}x_jt_jy_jy_{j+1}y_{j+2}y'_{j+3}$ is a subpath of $\Gamma_1$
and the path $R_2 =
z'_{j-1}z_jz_{j+1}t_{j+1}x_{j+1}x_{j+2}t_{j+2}z_{j+2}z'_{j+3}$ is a
subpath of $\Gamma_2$. See to the left part of Figure
\ref{Figure:local-transformation-3}.

Let us perform the following local transformation: delete $t_jz_j$,
$x_jx_{j+1}$, $t_{j+1}y_{j+1}$, $z_{j+1}z_{j+2}$ and
$t_{j+2}y_{j+2}$ from $M$ and add $x_jt_{j}$, $z_{j}z_{j+1}$,
$t_{j+1}x_{j+1}$, $y_{j+1}y_{j+2}$ and $t_{j+2}z_{j+2}$. Let $M'$ be
the resulting perfect matching. Then the subpath $R_1$ of $\Gamma_1$
is replaced by $R'_1 =
x'_{j-1}x_jx_{j+1}x_{j+2}t_{j+2}y_{j+2}y'_{j+3}$ and the subpath
$R_2$ of $\Gamma_2$ is replaced by $R'_2 =
z'_{j-1}z_jt_{j}y_{j}y_{j+1}t_{j+1}z_{j+1}z_{j+2}z'_{j+3}$. We
obtain a new $2$-factor containing two new cycles named
$\Gamma'_{1}$ and $\Gamma'_{2}$ such that $l(\Gamma'_1) =
l(\Gamma_1)$ and $\ l(\Gamma'_2) = l(\Gamma_2)$ (see Figure
\ref{Figure:local-transformation-3}). It is clear that for $i \in
{\bf Z}_{k}^{} \setminus \{j,j+1,j+2\}$ $C_i$ is $\Gamma'_2$-major
(respectively $\Gamma'_1$-major) if and only if $C_i$ is
$\Gamma_2$-major (respectively $\Gamma_1$-major). Note that $C_j$
and $C_{j+1}$ are $\Gamma'_2$-major and $C_{j+2}$ is
$\Gamma'_1$-major.
\end{Prf}

\vspace{1cm}

The operation depicted in Lemma \ref{Lemma:Gamma1-Gamma2-Gamma2}
above will be called a {\em local transformation of type 3}.

 \vspace{5mm}
\begin{Lem} \label{Lemma:2_factor_type1_3}
Let $M$ be a perfect matching of type $1$ of $\ G=FS(j,k)$ such that
the $2-$factor $G \setminus M$ has exactly two cycles $\Gamma_1$ and
$\Gamma_2$ such that $l(\Gamma_1) \leq l(\Gamma_2)$ and
$l(\Gamma_2)$ is as great as possible. Then there exists at most one
$\Gamma_1$-major claw.
\end{Lem}

\begin{Prf}
Suppose, for the sake of contradiction, that there exist at least
two $\Gamma_1$-major claws. Since $l(\Gamma_2)$ is maximum, by Lemma
\ref{Lemma:Gamma1-Gamma1} these claws are not consecutive. Then
consider two $\Gamma_1$-major claws $C_i$ and $C_{i+h+1}$ (with $h
\geq 1$) such that the $h$ consecutive claws $(C_{i+1}, \ldots,
C_{i+h})$ are $\Gamma_2$-major. Since $l(\Gamma_2)$ is maximum, by
Lemma \ref{Lemma:Gamma1-Gamma2-Gamma1} the number $h$ is at least
$2$. Then by applying $r = \lfloor \frac h 2 \rfloor$ consecutive
local transformations of type $3$ (Lemma
\ref{Lemma:Gamma1-Gamma2-Gamma2}) we obtain a perfect matching
$M^{(r)}$ such that the $2-$factor $G \setminus M^{(r)}$ has exactly
two cycles $\Gamma^{(r)}_1$ and $\Gamma^{(r)}_2$ with
$l(\Gamma^{(r)}_1) = l(\Gamma_1)$ and $l(\Gamma^{(r)}_2) =
l(\Gamma_2)$ and such that $C_{i+2\lfloor \frac h 2 \rfloor}$ and
$C_{i+h+1}$ are $\Gamma^{(r)}_1$-major. Since $l(\Gamma^{(r)}_2)$ is
maximum, we can conclude by Lemma \ref{Lemma:Gamma1-Gamma1} and by
Lemma \ref{Lemma:Gamma1-Gamma2-Gamma1} that $h$ is neither even nor
odd, a contradiction.
\end{Prf}

\subsection{Perfect matchings of type 2}
\label{subsect:type-2}

We give here a structural result about perfect matchings of type 2
in $\ G=FS(j,k)$.

\begin{Lem} \label{Lemma:2_factor_type2_1}
Let $M$ be a perfect matching of type $2$ of $G=FS(j,k)$ (with $k
\geq 4$). Then the $2-$factor $G \setminus M$ has exactly one cycle
of even length $l \geq k$ and a set of $p$ cycles of length $6$
where $l+6p=4k$ (with $0 \leq p \leq {\frac k 2}$).
\end{Lem}

\begin{Prf}
Let $M$ be a perfect matching of type $2$ in $G$. By Lemma
\ref{Lemma:PerfectMatching} the number $k$ of claws is even. Let $i$
in ${\bf Z}_{k}^{}$ such that there are two edges of $M$ between
$C_{i-1}$ and $C_{i}$. There are no edges of $M$ between $C_{i}$ and
$C_{i+1}$ and two edges of $M$ between $C_{i+1}$ and $C_{i+2}$. We
may consider that $ \ 0 \leq i < k-1$.\\
For $j\in \{i, i+2, i+4, \ldots\}$ we denote by $e_j$ the unique
edge of $G \setminus M$ having one end vertex in $C_{j-1}$ and the
other in $C_{j}$. Let us denote by $A$ the set $\{e_{i},
e_{i+2},e_{i+4},\ldots\}$. We note that $\mid A \mid = {\frac k 2}$.\\
Assume without loss of generality that the two edges of $M$ between
$C_{i-1}$ and $C_{i}$ have end vertices in $C_{i}$ which are $x_{i}$
and $y_{i}$ (then $z_i$ is the end vertex of $e_i$ in $C_i$). Two
cases may now occur.

{\bf Case 1:} {\em The end vertices in $C_{i+1}$ of the two edges of
$M$ between $C_{i+1}$ and $C_{i+2}$ are $x_{i+1}$ and $y_{i+1}$}
(then $z_{i+1}$ is the end vertex of $e_{i+2}$ in $C_{i+1}$).\\
In that case the $2-$ factor $G \setminus M$ contains the cycle of
length $6$ $x_{i}x_{i+1}t_{i+1}y_{i+1}y_{i}t_{i}$ while the edge
$z_{i}z_{i+1}$ of $G \setminus M$ relies $e_i$ and $e_{i+2}$.

{\bf Case 2:} {\em The end vertices in $C_{i+1}$ of the two edges of
$M$ between $C_{i+1}$ and $C_{i+2}$ are $y_{i+1}$ and $z_{i+1}$
(respectively  $x_{i+1}$ and $z_{i+1}$)}. Then $x_{i+1}$
(respectively $y_{i+1}$) is the end vertex of $e_{i+2}$ in
$C_{i+1}$.\\
In that case the edges $e_i$ and $e_{i+2}$ are connected in $G
\setminus M$ by the path
$z_{i}z_{i+1}t_{i+1}y_{i+1}y_{i}t_{i}x_{i}x_{i+1}$ (respectively
$z_{i}z_{i+1}t_{i+1}x_{i+1}x_{i}t_{i}y_{i}y_{i+1}$).

The same reasoning can be done for $\{e_{i+2}, e_{i+4}\}$,
$\{e_{i+4}, e_{i+6}\}$, and so on. Then, we see that the set $A$ is
contained in a unique cycle $\Gamma$ of $G \setminus M$ which
crosses each claw. Thus, the length $l$ of $\Gamma$ is at least $k$.
More precisely, each $e_j$ in $A$ contributes for $1$ in $l$, in
Case 1 the edge $\ z_{i}z_{i+1}$ contributes for $1$ in $l$ and in
Case 2 the path $z_{i}z_{i+1}t_{i+1}y_{i+1}y_{i}t_{i}x_{i}x_{i+1}$
contributes for $7$ in $l$. Let us suppose that Case 1 appears $p$
times ($0 \leq p \leq {\frac k 2}$), that is to say $G \setminus M$
contains $p$ cycles of length $6$. Since Case 2 appears ${\frac k 2}
-p$ times, the length of $\Gamma$ is  $l = {\frac k 2} + p +
7({\frac k 2} -p) = 4k -6p$.
\end{Prf}

\begin{Rem} \label{Remark:even 2-factor}
If $k$ is even then by Lemmas \ref{Lemma:2_factor_type1_1},
\ref{Lemma:2_factor_type1_2} and \ref{Lemma:2_factor_type2_1} $\
FS(j,k)$ has an even $2$-factor. That is to say $\ FS(j,k)$ is a
\cubthree.
\end{Rem}

\section{Perfect matchings and hamiltonian cycles of $F(j,k)$}

\subsection{Perfect matchings of type 1 and hamiltonicity}
\label{subsect:type-1-hamilton}

\begin{Thm} \label{Theorem:Hamiltonian_1}
Let $M$ be a perfect matching of type $1$ of $\ G = FS(j,k)$. Then
the $2$-factor $G \setminus M$ is a hamiltonian cycle except for $k$
odd and $j=2$, and for $k$ even and $j=1\ {\rm or}\ 3$.
\end{Thm}

\begin{Prf}
Suppose that there exists a perfect matching $M$ of type $1$ of $G$
such that $G \setminus M$ is not a hamiltonian cycle. By Lemma
\ref{Lemma:2_factor_type1_1} and Lemma \ref{Lemma:2_factor_type1_2}
the 2-factor $G \setminus M$ is made of exactly two cycles
$\Gamma_{1}$ and $\Gamma_{2}$ whose lengths have the same parity as
$k$. Without loss of generality we suppose that $l(\Gamma_1) \leq
l(\Gamma_2)$. Assume moreover that among the perfect matchings of
type 1 of $G$ such that the $2-$factor $G \setminus M$ is composed
of two cycles, $M$ has been chosen in such a way that the length of
the longest cycle $\Gamma_2$ is as great as possible. By Lemma
\ref{Lemma:2_factor_type1_3} there exists at most one
$\Gamma_1$-major claw.

{\bf Case 1:} There exists one $\Gamma_1$-major claw.

Without loss of generality, suppose that $C_{0}$ is intersected by
$\Gamma_1$ in $\{y_0, t_0, x_0\}$ and that $y'_{k-1}y_0$ belongs to
$\Gamma_1$. Since for every $i \neq 0$ the claw $C_i$ is
$\Gamma_2$-major, $\Gamma_1$ contains the vertices $y_0, t_0,x_0,
x_1, x_2, \ldots, x_{k-1}$. \

$\bullet$ If $k = 2r+1$ with $r \geq 1$ then $\Gamma_2$ contains the path\\
 $$z_0 z_1 t_1 y_1 y_2 t_2 z_2... z_{2r-1}
t_{2r-1} y_{2r-1} y_{2r} t_{2r} z_{2r}.$$ \\Thus, $y_0x_{k-1}$,
$x_0y_{k-1}$, $z_0z_{k-1}$ are edges of $G$. This means that
$\cup_{i=0}^{i=k-1}\{C_i \setminus \{t_i\} \}$ induces two cycles,
that is to say $j=2$ and $G = FS(2,k)$. \

$\bullet$ If $k = 2r+2$ with $r \geq 1$ then $\Gamma_2$ contains the path\\
 $$z_0 z_1 t_1 y_1 y_2 t_2 z_2... z_{2r-1}
t_{2r-1} y_{2r-1} y_{2r} t_{2r} z_{2r} z_{2r+1} t_{2r+1} y_{2r+1}.$$
Thus, $x_0z_{k-1}$, $y_0x_{k-1}$ and $z_0y_{k-1}$ are edges. This
means that $\cup_{i=0}^{i=k-1}\{C_i \setminus \{t_i\} \}$ induces
one cycle, that is to say $j=1$ and $G = FS(1,k)$.

{\bf Case 2:} There is no $\Gamma_1$-major claw.

Suppose that $x_0$ belongs to $\Gamma_1$. Then, $\Gamma_1$ contains
$x_0, x_1,..., x_{k-1}$. \

$\bullet$ If $k= 2r+1$ with $r \geq 1$ then $\Gamma_2$ contains the
path \
 $$y_0 t_0 z_0 z_1 t_1 y_1 y_2...  z_{2r-1}
t_{2r-1} y_{2r-1} y_{2r} t_{2r} z_{2r}.$$
 Thus, $x_0 x_{k-1}$, $y_0
z_{k-1}$ and $z_0 y_{k-1}$ are edges of $G$ and the set
$\cup_{i=0}^{i=k-1}\{C_i \setminus \{t_i\} \}$ induces two
cycles,that is to say $j=2$ and $G = FS(2,k)$. \

$\bullet$ If $k= 2r+2$ with $r \geq 1$ then $\Gamma_2$ contains the
path \
 $$y_0 t_0 z_0 z_1 t_1 y_1 y_2... y_{2r} t_{2r}
z_{2r} z_{2r+1} t_{2r+1} y_{2r+1}.$$ Thus, $x_0x_{k-1}$,
$y_0y_{k-1}$ and $z_0z_{k-1}$ are edges. This means that
$\cup_{i=0}^{i=k-1}\{C_i \setminus \{t_i\} \}$ induces three cycles,
that is to say $j=3$ and $G = FS(3,k)$.

\end{Prf}

\begin{Def} \label{Definition:2-factor-Hamiltonian}
A cubic graph $G$ is said to be {\em $2$-factor hamiltonian}
\cite{FJLS-2003} if every $2$-factor of $G$ is a hamiltonian cycle
(or equivalently, if for every perfect matching $M$ of $G$ the
$2$-factor $G \setminus M$ is a hamiltonian cycle).
\end{Def}

By Theorem \ref{Theorem:Hamiltonian_1} for any odd $k$ $\geq 3$ and
$j\in \{1,3\}$ or for any even $k$ and $j=2$, and for every perfect
matching $M$ of type 1 in $\ FS(j,k)$ the $2$-factor $\ FS(j,k)
\setminus M$ is a hamiltonian cycle. By Lemma
\ref{Lemma:2_factor_type2_1} $\ FS(2,k)$ ($k \geq 4$) may have a
perfect matching $M$ of type 2 such that the $2$-factor $FS(2,k)
\setminus M$ is not a hamiltonian cycle (it may contains cycles of
length $6$).

Then we have the following.

\begin{Cor} \label{Corollary:Hamiltonian_2}
A graph $G= FS(j,k)$ is $2$-factor hamiltonian if and only if $k$ is
odd and $j=1\ {\rm or}\ 3$.
\end{Cor}

We note that $FS(1,3)$ is the "Triplex Graph" of Robertson, Seymour
and Thomas \cite{RobSey}. We shall examine others known results
about $2$-factor hamiltonian cubic graphs in Section
\ref{Section:2-factor-hamiltonian}.

\begin{Cor} \label{Corollary:FlowerSnark}
The chromatic index of a graph $G= FS(j,k)$ is $4$ if and only if
$j=2$ and $k$ is odd.
\end{Cor}

\begin{Prf}
When $j=2$ and $k$ is odd, any $2$-factor must have at least two
cycles, by Theorem \ref{Theorem:Hamiltonian_1}. Then Lemma
\ref{Lemma:2_factor_type1_2} implies that any $2$-factor is composed
of two odd cycles. Hence $G$ has chromatic index $4$.

When $j=1 \ {\rm or} \ 3$ and $k$ is odd by Theorem
\ref{Theorem:Hamiltonian_1} $FS(j,k)$ is hamiltonian. If $k$ is even
then by Lemmas \ref{Lemma:2_factor_type1_1},
\ref{Lemma:2_factor_type1_2} and \ref{Lemma:2_factor_type2_1} $\
FS(j,k)$ has an even $2$-factor.
\end{Prf}

\subsection{Perfect matchings of type 2 and hamiltonicity}
\label{subsect:type-2-hamilton}

At this point of the discourse one may ask what happens for perfect
matchings of type $2$ in $FS(j,k)$ ($k$ even). Can we characterize
and count perfect matchings of type $2$, complementary $2$-factor of
which is a hamiltonian cycle~? An affirmative answer shall be given.

Let us consider a perfect matching $M$ of type $2$ in $FS(j,2p)$
with $p \geq 2$. Suppose that there are no edges of $M$ between
$C_{2i-1}$ and $C_{2i}$ (for any $i \geq 1$), that is $M$ is a
matching of type $2.0$ (see Definition
\ref{Definition_type_of_matching}). Consider two consecutive claws
$C_{2i}$ and $C_{2i+1}$ ($0 \leq i \leq p-1$). There are three
cases:

Case (x): $\{y_{2i}y_{2i+1}, z_{2i}z_{2i+1}\} \subset M $  (then, $M
\cap (C_{2i}\cup C_{2i+1})= \{x_{2i}t_{2i}, x_{2i+1}t_{2i+1}\}$).

Case (y): $\{x_{2i}x_{2i+1}, z_{2i}z_{2i+1}\} \subset M $  (then, $M
\cap (C_{2i}\cup C_{2i+1})= \{y_{2i}t_{2i}, y_{2i+1}t_{2i+1}\}$).

Case (z): $\{x_{2i}x_{2i+1}, y_{2i}y_{2i+1}\} \subset M $  (then, $M
\cap (C_{2i}\cup C_{2i+1})= \{z_{2i}t_{2i}, z_{2i+1}t_{2i+1}\}$).

The subgraph induced on $C_{2i}\cup C_{2i+1}$ is called a {\em
block}. In Case (x) (respectively Case (y), Case (z)) a block is
called a {\em block of type $X$} (respectively {\em block of type
$Y$, block of type $Z$}). Then $FS(j,2p)$ with a perfect matchings
$M$ of type $2.0$ can be seen as a sequence of $p$ blocks properly
relied. In other words, a perfect matchings $M$ of type $2$ in
$FS(j,2p)$ is entirely described by a word of length $p$ on the
alphabet of three letters $\{ X, Y, Z\}$. The block $C_{0}\cup
C_{1}$ is called {\em initial block} and the block $C_{2p-1}\cup
C_{2p}$ is called {\em terminal block}. These extremal blocks are
not considered here as consecutive blocks.

By Lemma \ref{Lemma:2_factor_type2_1}, $FS(j,2p) \setminus M$ has no
$6$-cycles if and only if $FS(j,2p) \setminus M$ is a unique even
cycle. It is an easy matter to prove that two consecutive blocks do
not induce a $6$-cycle if and only if they are not of the same type.
Then the possible configurations for two consecutive blocks are
$XY$, $XZ$, $YX$, $YZ$, $ZX$ and $ZY$. To eliminate a possible
$6$-cycle in $C_{0}\cup C_{2p-1}$ we have to determine for every $j
\in \{1,2,3\}$ the forbidden extremal configurations. An extremal
configuration shall be denoted by a word on two letters in $\{ X, Y,
Z\}$ such that the left letter denotes the type of the initial block
$C_{0}\cup C_{1}$ and the right letter denotes the type of the
terminal block $C_{2p-1}\cup C_{2p}$. We suppose that the extremal
blocks are connected for $j=1$ by the edges $x_{2p-1}z_{0}$,
$y_{2p-1}x_{0}$ and $z_{2p-1}y_{0}$, for $j=2$ by the edges
$x_{2p-1}x_{0}$, $y_{2p-1}z_{0}$ and $z_{2p-1}y_{0}$ and for $j=3$
by the edges $x_{2p-1}x_{0}$, $y_{2p-1}y_{0}$ and $z_{2p-1}z_{0}$.
Then, it is easy to verify that we have the following result.

\begin{Lem} \label{Lemma:Forbidden-configurations}
Let $M$ be a perfect matching of type $2.0$ of $G=FS(j,2p)$ (with $p
\geq 2$) such that the $2-$factor $G \setminus M$ is a hamiltonian
cycle. Then the forbidden extremal configurations are

\hspace{1cm} $XY$, $YZ$ and $ZX$ for $FS(1,2p)$,

\hspace{1cm} $XX$, $YZ$ and $ZY$ for $FS(2,2p)$,

\hspace{1mm} and $\ XX$, $YY$ and $ZZ$ for $FS(3,2p)$.
\end{Lem}

Thus, any perfect matching $M$ of type $2.0$ of $FS(j,2p)$ such that
the $2-$factor $G \setminus M$ is a hamiltonian cycle is totally
characterized by a word of length $p$ on the alphabet $\{ X, Y, Z\}$
having no two identical consecutive letters and such that the
sub-word  [initial letter][terminal letter] is not a forbidden
configuration. Then, we are in position to obtain the number of such
perfect matchings in $FS(j,2p)$. Let us denote by $\mu'_{2.0}(j,2p)$
(respectively $\mu'_{2.1}(j,2p)$, $\mu'_{2}(j,2p)$) the number of
perfect matchings of type $2.0$ (respectively type $2.1$, type $2$)
complementary to a hamiltonian cycle in $FS(j,2p)$. Clearly
$\mu'_{2}(j,2p) = \mu'_{2.0}(j,2p) + \mu'_{2.1}(j,2p)$ and
$\mu'_{2.0}(j,2p) = \mu'_{2.1}(j,2p)$.

\begin{Thm} \label{Theorem:Hamiltonian_2} The numbers $\mu'_{2}(j,2p)$ of perfect
matchings of type $2$ complementary to hamiltonian cycles in
$FS(j,2p)$ ($j \in \{1, 2, 3\}$) are given by:

\hspace{1cm}  $\mu'_{2}(1,2p) = 2^{p+1} +(-1)^{p+1}2$,

\hspace{1cm}  $\mu'_{2}(2,2p) = 2^{p+1}\ $,

 and $\ \ \mu'_{2}(3,2p) = 2^{p+1} +(-1)^p 4$.
\end{Thm}

\begin{Prf}
Consider, as previously, perfect matchings of type $2.0$. Let
$\alpha$ and $\beta$ be two letters in $\{ X, Y, Z\}$ (not
necessarily distinct). Let $A^p_{\alpha\beta}$ be the set of words
of length $p$ on $\{ X, Y, Z\}$ having no two consecutive identical
letters, beginning by $\alpha$ and ending by a letter distinct from
$\beta$. Denote the number of words in $A^p_{\alpha\beta}$ by
$a^p_{\alpha\beta}$. Let $B^p_{\alpha\beta}$ be the set of words of
length $p$ on $\{ X, Y, Z\}$ having no two consecutive identical
letters, beginning by $\alpha$ and ending by $\beta$. Denote by
$b^p_{\alpha\beta}$ the number of words in $B^p_{\alpha\beta}$.

Clearly, the number of words of length $p$ having no two consecutive
identical letters and beginning by $\alpha$ is $2^{p-1}$. Then
$a^p_{\alpha\beta}+b^p_{\alpha\beta}= 2^{p-1}$. The deletion of the
last $\beta$ of a word in $B^p_{\alpha\beta}$ gives a word in
$A^{p-1}_{\alpha\beta}$ and the addition of $\beta$ to the right of
a word in $A^{p-1}_{\alpha\beta}$ gives a word in
$B^p_{\alpha\beta}$.

Thus $b^p_{\alpha\beta} = a^{p-1}_{\alpha\beta}$ and for every $p
\geq 3$ $a^{p}_{\alpha\beta} = 2^{p-1} - a^{p-1}_{\alpha\beta}$. We
note that $a^2_{\alpha\beta}=2$ if $\alpha = \beta$, and
$a^2_{\alpha\beta}=1$ if $\alpha \neq \beta$. If $\alpha = \beta$ we
have to solve the recurrent sequence : $u_2=2$ and $u_{p}= 2^{p-1} -
u_{p-1}$ for $p \geq 3$. If $\alpha \neq \beta$ we have to solve the
recurrent sequence : $v_2=1$ and $v_{p}= 2^{p-1} - v_{p-1}$ for $p
\geq 3$. Then we obtain $\ u_p = \frac 2 3 (2^{p-1} +(-1)^p)\ $ and
$\ v_p = \frac 1 3 (2^{p} +(-1)^{p+1})$ for $p \geq 2$.

By Lemma \ref{Lemma:Forbidden-configurations}

\hspace{1cm} $\mu'_{2.0}(1,2p)= a^p_{XY}+a^p_{YZ}+a^p_{ZX} = 3 v_p =
2^{p} +(-1)^{p+1}$,

\hspace{1cm} $\mu'_{2.0}(2,2p)= a^p_{XX}+a^p_{YZ}+a^p_{ZY} =
u_p+2v_p =  2^{p}\ $,

and $\ \ \mu'_{2.0}(3,2p)= a^p_{XX}+a^p_{YY}+a^p_{ZZ} = 3 u_p =
2^{p} +(-1)^p 2$.

Since $\mu'_{2}(j,2p) = \mu'_{2.0}(j,2p) + \mu'_{2.1}(j,2p)$ and
$\mu'_{2.0}(j,2p) = \mu'_{2.1}(j,2p)$ we obtain the announced
results.
\end{Prf}

\begin{Rem} \label{Remark:Hamilton_2} We see that $\mu'_{2}(j,2p)
\simeq 2^{p+1}$ and this is to compare with the number
$\mu_{2}(j,2p) = 2\times 3^{p}$ of perfect matchings of type $2$ in
$FS(j,2p)$ (see backward in Section
\ref{Section:CountingPerfectMatchings}).
\end{Rem}

\subsection{Strong matchings and \Jaes}
\label{subsect-Jaeger-graph} For a given graph $G=(V,E)$ a {\em
strong matching} (or {\em induced matching}) is a matching $S$ such
that no two edges of $S$ are joined by an edge of $G$. That is, $S$
is the set of edges of the subgraph of $G$ induced by the set
$V(S)$. We consider cubic graphs having a perfect matching which is
the union of two strong matchings that we call {\em \Jae} (in his
thesis \cite{Jae} Jaeger called these cubic graphs {\em equitable}).
We call {\em Jaeger's matching} a perfect matching $M$ of a cubic
graph $G$ which is the union of two strong matchings $M_B$ and
$M_R$. Set $B =V(M_B)$ (the blue vertices) and $R= V(M_R)$ (the red
vertices). An edge of $G$ is said {\em mixed} if its end vertices
have distinct colours. Since the set of mixed edges is $E(G)
\setminus M$, the $2$-factor $G \setminus M$ is even and $\mid B
\mid \ =\  \mid M \mid$. Thus, every \Jae $G$ is a \cubthree and for
any Jaeger's matching $M = M_B \cup M_R $, $\mid M_B \mid \ =\ \mid
M_R\mid$. See, for instance, \cite{FTVW1} and \cite{FTVW2} for some
properties of these graphs.

In this subsection we determine the values of $j$ and $k$ for which
a graph $FS(j,k)$ is a \Jaev.

\begin{Lem} \label{Lemma:JaegerMatchingHasType1}
If $G=FS(j,k)$ is a \Jaev $\ {\rm(with}\ k \geq 3)$ and $M = M_B
\cup M_R $ is a Jaeger's matching of $G$ then $M$ is a perfect
matching of type $1$.
\end{Lem}

\begin{Prf}
Suppose that $M$ is of type 2 and suppose without loss of generality
that there are two edges of $M$ between $C_0$ and $C_1$, for
instance $x_0x_1$ and $y_0y_1$. Then $C_0 \cap M = \{t_0z_0\}$ and
$C_1\cap M = \{t_1z_1\}$. Suppose that $x_0x_1$ and $y_0y_1$ belong
to $M_B$. Since $M_B$ is a strong matching, $t_0z_0$ and $t_1z_1$
belong to $M \setminus M_B = M_R$. This is impossible because $M_R$
is also a strong matching. By symmetry there are no two edges of
$M_R$ between $C_0$ and $C_1$. Then there is one edge of $M_B$
between $C_0$ and $C_1$, $x_0x_1$ for instance, and one edge of
$M_R$ between $C_0$ and $C_1$, $y_0y_1$ for instance. Since $M_B$
and $M_R$ are strong matchings, there is no edge of $M$ in $C_0 \cup
C_1$, a contradiction. Thus, $M$ is a perfect matching of type 1.
\end{Prf}

\begin{Lem} \label{Lemma:JaegerGraphNecCond}
If $G=FS(j,k)$ is a \Jaev $\ {\rm(with}\ k \geq 3)$ then either
($j=1$ and $k \equiv 1\ {\rm or\ }2 \ (mod\ 3)$) or ($j=3$ and $k
\equiv 0 \ (mod\ 3)$).
\end{Lem}

\begin{Prf}
Let $M = M_B \cup M_R $ be a Jaeger's matching of $G$. By Lemma
\ref{Lemma:JaegerMatchingHasType1} $M$ is a perfect matching of type
$1$. Suppose without loss of generality that $M_B \cap E(C_0) =
\{x_0t_0\}$. Since $M_B$ is a strong matching there is no edge of
$M_B$ between $C_0$ and $C_1$. Suppose, without loss of generality,
that the edge in $M_R$ joining $C_0$ to $C_1$ is $y_0y_1$. Consider
the claws $C_0$, $C_1$ and $C_2$. Since $M_B$ and $M_R$ are strong
matchings, we can see that the choices of $x_0t_0 \in M_B$ and
$y_0y_1 \in M_R$ fixes the positions of the other edges of $M_B$ and
$M_R$. More precisely, $\{t_1z_1, y_2t_2\} \subset M_B$ and
$\{x_1x_2, z_2z'_3\} \subset M_R$. This unique configuration is
depicted in Figure \ref{Figure:JaegerGraph}.

\vspace{5mm}
\begin{figure}[htb]
\centering \epsfsize=0.6 \hsize \noindent \epsfbox{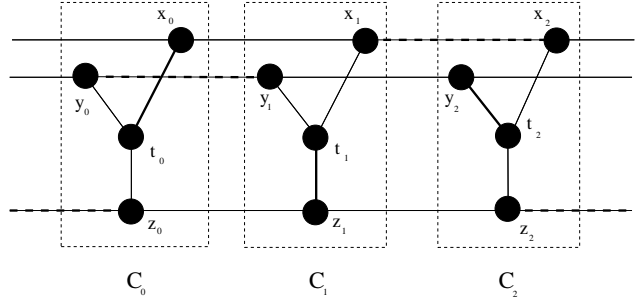}
\caption{Strong matchings $M_B$ (bold edges) and $M_R$ (dashed
edges)} \label{Figure:JaegerGraph}
\end{figure}

If $k \geq 4$ then we see that $z_2z_3 \in M_R$, $x_3t_3 \in M_B$,
and $y_3y'_4 \in M_R$. So, the local situation in $C_3$ is similar
to that in $C_0$, and we can see that there is a unique Jaeger's
matching $M = M_B \cup M_R$ such that $x_0t_0 \in M_B$ and $y_0y_1
\in M_R$ in the graph $FS(j,k)$. We have to verify the coherence of
the connections between the claws $C_{k-1}$ and $C_0$. We note that
$M_B = M \cap (\cup_{i=0}^{i=k-1} E(C_i))$ and $M_R$ is a strong
matching included in the $2$-factor induced by
$\cup_{i=0}^{i=k-1}\{V(C_i) \setminus \{t_i\} \}$.

{\bf Case 1:} $k = 3p$ with $p \geq 1$.\\
We have $x_0t_0 \in M_B$, $y_{k-1}t_{k-1} \in M_B$, $x_{k-2}x_{k-1}
\in M_R$ and $z'_{k-1}z_0 = z_{k-1}z'_0 \in M_R$ (that is,
$z_{k-1}z_0\in M_R$). Thus, $z_{k-1}z_0$, $y_{k-1}y_0$ and
$x_{k-1}x_0$ are edges of $FS(j, 3p)$ and we must have $j=3$.

{\bf Case 2:} $k = 3p+1$ with $p \geq 1$.\\
We have $x_0t_0 \in M_B$, $x_{k-1}t_{k-1} \in M_B$ (that is,
$x_{k-1}x_0 \notin E(G)$), $z_{k-2}z_{k-1} \in M_R$ and $z'_{k-1}z_0
= y_{k-1}y'_0 \in M_R$ (that is, $y_{k-1}z_0 \in M_R$). Thus,
$y_{k-1}z_0$, $x_{k-1}y_0$ and $z_{k-1}x_0$ are edges of $FS(j,
3p+1)$ and we must have $j=1$.

{\bf Case 3:} $k = 3p+2$ with $p \geq 1$.\\
We have $x_0t_0 \in M_B$, $z_{k-1}t_{k-1} \in M_B$, $y_{k-2}y_{k-1}
\in M_R$ and $z'_{k-1}z_0 = x_{k-1}x'_0 \in M_R$ (that is
$x_{k-1}z_0 \in M_R$). Thus, $x_{k-1}z_0$, $y_{k-1}x_0$ and
$z_{k-1}y_0$ are edges of $FS(j, 3p+2)$ and we must have $j=1$.
\end{Prf}

\begin{Rem} \label{Remark:FS(2,k)NotJaeger}
It follows from Lemma \ref{Lemma:JaegerGraphNecCond} that for every
$k \geq 3$ the graph $FS(2,k)$ is not a \Jaev. This is obvious when
$k$ is odd, since the flower snarks have chromatic index $4$.
\end{Rem}

Then, we obtain the following.

\begin{Thm} \label{Theorem:JaegerGraph}
For $j\in \{1, 2, 3\}$ and $k \geq 2$, the graph  $\ G = FS(j,k)$ is
a \Jae if and only if

\hspace{1cm} either  $\ k \equiv 1\ {\rm or\ }2 \ (mod\ 3)$ and
$j=1$,

\hspace{1cm} or  $\ k \equiv 0 \ (mod\ 3)$ and $j=3$.

Moreover, $FS(1,2)$ has $3$ Jaeger's matchings and for $k \geq 3$ a
\Jae $\ G= FS(j,k)$ has exactly $6$ Jaeger's matchings.
\end{Thm}

\begin{Prf}
For $k=2$ we remark that $FS(1,2)$ (that is the cube) has exactly
three distinct Jaeger's matchings $M_1$, $M_2$ and $M_3$. Following
our notations: $M_1 = \{x_0t_0, t_1z_1\} \cup \{y_0y_1, z_0x_1\}$,
$M_2 = \{z_0t_0, t_1y_1\} \cup \{y_0z_1, x_0x_1\}$ and $M_3 =
\{y_0t_0, t_1x_1\} \cup \{z_0z_1, x_0y_1\}$.

For $k \geq 3$, by  Lemma \ref{Lemma:JaegerGraphNecCond}, condition
$$ (*) \ (j=1\ {\rm and}\ k \equiv 1\ {\rm or\ }2 \ (mod\ 3))\ {\rm
or}\ (j=3\ {\rm and}\  k \equiv 0 \ (mod\ 3))$$ is a necessary
condition for $\ FS(j,k)$ to be a \Jaev.

Consider the function $\Phi_{X,Y} : V(G) \rightarrow V(G)$ such that
for every $i$ in ${\bf Z}_{k}^{}$, $\Phi_{X,Y}(t_i)=t_i$,
$\Phi_{X,Y}(z_i)=z_i$, $\Phi_{X,Y}(x_i)=y_i$ and
$\Phi_{X,Y}(y_i)=x_i$. Define similarly $\Phi_{X,Z}$ and
$\Phi_{Y,Z}$. For $j= 1 \ {\rm or} \ 3$ these functions are
automorphisms of $FS(j,k)$. Thus, the process described in the proof
of Lemma \ref{Lemma:JaegerGraphNecCond} is a constructive process of
all Jaeger's matchings in a graph $FS(j,k)$ (with $k \geq 3$)
verifying condition (*).

We remark that for any choice of an edge $e$ of $C_0$ to be in $M_B$
there are two distinct possible choices for an edge $f$ between
$C_0$ and $C_1$ to be in $M_R$, and such a pair $\{e, f\}$
corresponds exactly to one Jaeger's matching. Then, a \Jae $FS(j,k)$
(with $k \geq 3$) has exactly $6$ Jaeger's matchings.
\end{Prf}

\begin{Rem} \label{Remark:BergeFulkersonConjecture}
The {\em Berge-Fulkerson Conjecture} states that if $G$ is a
bridgeless cubic graph, then there exist six perfect matchings $
M_1,\ldots,M_6 $ of $G$ (not necessarily distinct) with the property
that every edge of $G$ is contained in exactly two of
$M_1,\ldots,M_6$ (this conjecture is attributed to Berge in
\cite{Sey} but appears in \cite{Fulk}). Using each colour of a
\cubthree twice, we see that such a graph verifies the
Berge-Fulkerson Conjecture. Very few is known about this conjecture
except that it holds for the Petersen graph and for \cubthreesv. So,
Berge-Fulkerson Conjecture holds for \Jaesv, but generally we do not
know if we can find six distinct perfect matchings. We remark that
if $FS(j,k)$, with $k \geq 3$, is a \Jae then its six Jaeger's
matchings are such that every edge is contained in exactly two of
them.
\end{Rem}

\section{$2$-factor hamiltonian cubic graphs}
\label{Section:2-factor-hamiltonian} Recall that a simple graph of
maximum degree $d>1$ with edge chromatic number equal to $d$ is said
to be a {\em Class $1$ graph}. For any $d$-regular simple graph
(with $d>1$) of even order and of Class $1$, for any minimum
edge-colouring of such a graph, the set of edges having a given
colour is a perfect matching (or $1$-factor). Such a regular graph
is also called a {\em $1$-factorable graph}. A Class $1$ $d$-regular
graph of even order is {\em strongly hamiltonian} or {\em perfectly
$1$-factorable} (or is a {\em Hamilton graph} in the Kotzig's
terminology \cite{Kotz}) if it has an edge colouring such that the
union of any two colours is a hamiltonian cycle. Such an edge
colouring is said to be {\em a Hamilton decomposition} in the
Kotzig's terminology. In \cite{KotzRussian} by using two operations
$\rho$ and $\pi$ (described also in \cite{Kotz}) and starting from
the $\theta$-graph (two vertices joined by three parallel edges) he
obtains all strongly hamiltonian cubic graphs, but these operations
does not always preserve planarity. In his paper \cite{Kotz} he
describes a method for constructing planar strongly hamiltonian
cubic graphs and he deals with the relation between strongly
hamiltonian cubic graphs and $4$-regular graphs which can be
decomposed into two hamiltonian cycles. See also \cite{KotzLab} and
a recent work on strongly hamiltonian cubic graphs \cite{BonMaz} in
which the authors give a new construction of strongly hamiltonian
graphs.

A Class $1$ regular graph such that every edge colouring is a
Hamilton decomposition is called a {\em pure Hamilton graph} by
Kotzig \cite{Kotz}. Note that $K_4$ is a pure Hamilton graph and
every cubic graph obtained from $K_4$ by a sequence of triangular
extensions is also a pure Hamilton cubic graph. In the paper
\cite{Kotz} of Kotzig, a consequence of his Theorem 9 (p.77)
concerning pure Hamilton graphs is that the family of pure Hamilton
graphs that he exhibits is precisely the family obtained from $K_4$
by triangular extensions. Are there others pure Hamilton cubic
graphs~? The answer is "yes".

We remark that $2$-factor hamiltonian cubic graphs defined above
(see Definition \ref{Definition:2-factor-Hamiltonian}) are pure
Hamilton graphs (in the Kotzig's sense) but the converse is false
because $K_4$ is $2$-factor hamiltonian and the pure Hamilton cubic
graph on $6$ vertices obtained from $K_4$ by a triangular extension
(denoted by $PR_3$) is not $2$-factor hamiltonian. Observe that the
operation of triangular extension preserves the property "pure
Hamilton", but does not preserve the property "$2$-factor
hamiltonian". The Heawood graph $H_0$ (on $14$ vertices) is pure
Hamiltonian, more precisely it is $2$-factor hamiltonian (see
\cite{Funk-Labbate} Proposition 1.1 and Remark 2.7). Then, the
graphs obtained from the Heawood graph $H_0$ by triangular
extensions are also pure Hamilton graphs.

A {\em minimally $1$-factorable} graph $G$ is defined by Labbate and
Funk \cite{Funk-Labbate} as a Class $1$ regular graph of even order
such that every perfect matching of $G$ is contained in exactly one
$1$-factorization of $G$. In their article they study bipartite
minimally $1$-factorable graphs and prove that such a graph $G$ has
necessarily a degree $d \leq 3$. If $G$ is a minimally
$1$-factorable cubic graph then the complementary $2$-factor of any
perfect matching has a unique decomposition into two perfect
matchings, therefore this $2$-factor is a hamiltonian cycle of $G$,
that is $G$ is $2$-factor hamiltonian. Conversely it is easy to see
that any $2$-factor hamiltonian cubic graph is minimally
$1$-factorable.
%Then, a cubic graph is minimally $1$-factorable if
%and only if it is $2$-factor hamiltonian.
The complete bipartite graph $K_{3,3}$ and the Heawood graph $H_0$
are examples of $2$-factor hamiltonian bipartite graph given by
Labbate and Funk. Starting from $H_0$, from $K_{1,3}$ and from three
copies of any tree of maximum degree $3$ and using three operations
called {\em amalgamations} the authors exhibit an infinite family of
bipartite $2$-factor hamiltonian cubic graphs, namely the
$poly-HB-R-R^2$ graphs (see \cite{Funk-Labbate} for more details).
Except $H_0$, these graphs are exactly cyclically $3$-edge
connected. Others structural results about $2$-factor hamiltonian
bipartite cubic graph are obtained in \cite{Labbate-2001},
\cite{Labbate-2002}. These results have been completed and a simple
method to generate $2$-factor hamiltonian bipartite cubic graphs was
given in \cite{FJLS-2003}.

\begin{Prop} {\rm (Lemma 3.3, \cite{FJLS-2003})}
Let $G$ be a $2$-factor hamiltonian bipartite cubic graph. Then $G$
is $3$-connected and $\mid V(G) \mid \equiv 2 \ (mod\ 4)$.
\end{Prop}

Let $G_1$ and $G_2$ be disjoint cubic graphs, $x \in v(G_1)$,  $y
\in v(G_2)$. Let $x_1, x_2, x_3$ (respectively $y_1, y_2, y_3$) be
the neighbours of $x$ in $G_1$ (respectively, of $y$ in $G_2$). The
cubic graph $G$ such that $V(G) = (V(G_1) \setminus \{x\}) \cup
(V(G_2) \setminus \{y\})$ and $E(G) = (E(G_1) \setminus \{x_1x,
x_2x, x_3x\}) \cup (E(G_2) \setminus \{y_1y, y_2y, y_3y\}) \cup
\{x_1y_1, x_2y_2, x_3y_3\}$ is said to be a {\em star product} and
$G$ is denoted by $(G_1,x)*(G_2,y)$. Since $\{x_1y_1, x_2y_2,
x_3y_3\}$ is a cyclic edge-cut of $G$, a star product of two
$3$-connected cubic graphs has cyclic edge-connectivity $3$.

\begin{Prop} {\rm (Proposition 3.1, \cite{FJLS-2003})}
If a bipartite cubic graph $G$ can be represented as a star product
$G = (G_1,x)*(G_2,y)$, then $G$ is $2$-factor hamiltonian if and
only if $G_1$ and $G_2$ are $2$-factor hamiltonian.
\end{Prop}

Then, taking iterated star products of $K_{3,3}$ and the Heawood
graph $H_0$ an infinite family of $2$-factor hamiltonian cubic
graphs is obtained. These graphs (except $K_{3,3}$ and $H_0$) are
exactly cyclically $3$-edge connected. In \cite{FJLS-2003} the
authors conjecture that the process is complete.

\begin{Conj} \label{Conjecture:FJLS-2003}
({\rm Funk, Jackson, Labbate, Sheehan (2003)\cite{FJLS-2003}}) Let
$G$ be a bipartite $2$-factor hamiltonian cubic graph. Then $G$ can
be obtained from $K_{3,3}$ and the Heawood graph $H_0$ by repeated
star products.
\end{Conj}

The authors precise that a smallest counterexample to Conjecture
\ref{Conjecture:FJLS-2003} is a cyclically $4$-edge connected cubic
graph of girth at least $6$, and that to show this result it would
suffice to prove that $H_0$ is the only $2$-factor hamiltonian
cyclically $4$-edge connected bipartite cubic graph of girth at
least $6$. Note that some results have been generalized  in
\cite{ADJLS-2008}.

To conclude, we may ask what happens for non bipartite $2$-factor
hamiltonian cubic graphs. Recall that $K_4$ and $FS(1,3)$ (the
"Triplex Graph" of Robertson, Seymour and Thomas \cite{RobSey}) are $2$-factor
hamiltonian cubic graphs. By Corollary \ref{Corollary:Hamiltonian_2}
the graphs $FS(j,k)$ with $k$ odd and $j=1\ {\rm or}\ 3$ introduced
in this paper form a new infinite family of non bipartite $2$-factor
hamiltonian cubic graphs. We remark that they are cyclically
$6$-edge connected. Can we generate others families of non bipartite
$2$-factor hamiltonian cubic graphs~? Since $PR_3$ (the cubic graph
on $6$ vertices obtained from $K_4$ by a triangular extension) is
not $2$-factor hamiltonian and $PR_3 = K_4*K_4$, the star product
operation is surely not a possible tool.
\bibliographystyle{plain}

\bibliography{Bibliographie}

\end{document}